\documentstyle[epsfig]{aipproc}

\def\etal   {{et~al.}\ }

\def\vol#1  {{{#1}{\rm,}\ }}
\def\lya{{\rm Ly}\alpha}

\def\etal{et al.\ }

\def\cf{{cf.}\ }

\newcount\refno
\newcount\rfno
\def\eq{$^{\the\refno\ }$\advance\refno by 1}
\def\ad{\advance\rfno by 1}

\def\clock{\count0=\time \divide\count0 by 60
     \count1=\count0 \multiply\count1 by -60 \advance\count1 by \time
     \number\count0:\ifnum\count1<10{0\number\count1}\else\number\count1\fi}

\begin{document}
\title{The Intergalactic Medium \\ and Soft X-ray Background}

\author{Renyue Cen$^*$}
\address{$^*$Princeton University Observatory\thanks{Present research
is sponsored in part by the National Science Foundation and the National
Aeronautics and Space Administration.}\\
$^{\dagger}$Princeton, NJ 08544}

\maketitle

\begin{abstract}
I present an overview of some
of the recent advances in our understanding
of the distribution and evolution of the ordinary, baryonic matter
in the universe. Two observations that
strongly suggest that most of the baryons seen at high
redshift ($z\ge 2$) have turned into some forms yet undetected at $z=0$
are highlighted.
With the aid of large-scale cosmological hydrodynamic simulations, 
it is shown that most of the baryons today are in a gaseous form 
with a temperature of $10^5-10^7~$Kelvin -- the ``warm/hot gas",
shock heated during the gravitational collapse and formation
of the large-scale structure at low redshift.
Primarily line emissions from this warm/hot gas may account for 
a large fraction of the residual 
(after removal of identifiable discrete sources)
soft X-ray background at $h\nu < 1.0$keV.
How this warm/hot gas may be detected by the next generation
of EUV and soft X-ray instruments is indicated.
Detection or non-detection of this warm/hot gas
will have profound implications for cosmology.

\end{abstract}

\section*{Introduction}
%
%
%
%

It is well known that
most of the matter in the universe is in some non-luminous, dark form,
with the cold dark matter being the most popular choice
(Peebles 1993).
Not only that, most of the ordinary, baryonic matter, which
altogether makes up about 10-20\% of the total matter
in the universe, seems to be missing in the present=day universe.
At redshift $z=2-3$,
the amount of gas contained in the Lyman alpha forest
is (Rauch \etal 1997; Weinberg \etal 1997)
\begin{equation}
\Omega_{b,\lya} (z=2-3) \ge 0.017h^{-2} = 0.035,
\end{equation}
\noindent 
where $\Omega_{b,\lya}(z=2-3)$ is 
the baryonic density in units of the critical density
extrapolated to $z=0$ and 
a Hubble constant $h\equiv \hbox{H}_0/(100$km/s/Mpc$)=0.70$ 
is adopted throughout.
Independently, the observed light-element ratios (in particular,
the deuterium to hydrogen ratio) in some carefully
selected absorption line systems at $z=2-3$,
interpreted within the context of the standard nucleosynthesis theory,
yield the total baryonic density (Burles \& Tytler 1998)
\begin{equation}
\Omega_{b,D/H} (z=2-3) = (0.019\pm 0.001)h^{-2} = 0.039\pm 0.002.
\end{equation}
\noindent 
The agreement between these two 
completely independent measurements is remarkable.
But, at redshift zero, after summing over all well observed contributions,
the baryonic density 
appears to be far (by a factor of three) below 
that indicated by equations (1) and (2)
(e.g., Fukugita, Hogan, \& Peebles 1997):
\begin{equation}
\Omega_{b}(z=0)|_{\hbox{seen}}=\Omega_{*} + \Omega_{HI} + \Omega_{H_2} + \Omega_{Xray,cl} \approx 0.0068 \le 0.011~~(2\sigma~\hbox{limit}).
\end{equation}
\noindent 
Thus, unless two independent errors
have been made in the arguments that led to 
equations (1) and (2),
there is a sharp decline of  the amount of observed baryons
from high redshift to the present day;
i.e., most of the baryons in the present-day universe
are yet to be detected.
Now, the evolution of universal baryons 
can be computed from standard initial 
conditions in the modern cosmological setting, 
using realistic large-scale hydrodynamic simulations.
It is found that the large amount of baryons are {\it not missing} but
{\it hidden} in an intergalactic gas
with a temperature $10^5<T<10^7~$Kelvin - the ``warm/hot gas" - at $z=0$,
which is difficult to detect (\S 2).
\S 3 shows observable signatures of the warm/hot gas
presents imprinted in the soft X-ray background.
Several ways to detect this warm/hot gas
are suggested in \S 4
and conclusions presented in \S 5.

\section*{Evolution of Cosmic Baryons}

In a series of model simulations
of nearly a dozen different models covering the current
interest we have shown,
consistently and robustly,
that from $50\%$ to $70\%$ of all baryons in all models examined
are shock heated during the gravitational collapse of the large-scale
structure in the recent past 
and are in a warm/hot gas with a temperature of $10^5-10^7~$Kelvin 
at $z=0$ (Ostriker \& Cen 1996)
with each model being approximately
normalized to match the local large-scale structure.
It was clearly and immdiately understood
that this model independent outcome found in earlier work
would have profound implications for cosmology and 
it is therefore of paramount importance  
to verify it with latest simulations that
include more relevant physics (including
feedback from star formation and
metal cooling, etc.)
and have more accurate treatment of shocks
(Ryu \etal 1993) than our prior code (Cen 1992).
Indeed, it was confirmed by a recent higher-resolution, 
larger-size hydrodynamic simulation of
a cold dark matter model with a cosmological constant.
The adopted model - the Ostriker-Steinhardt (1995) Concordance model -
is normalized to both the microwave background temperature
fluctuations measured by COBE 
on large scales (Bunn \& White 1997)
and the observed abundance of clusters of galaxies in
the local universe (Cen 1998)
with $\Omega_0=0.37$, $\Omega_{b}=0.049$,
$\Lambda_0=0.63$, $\sigma_8=0.80$,
$h=0.70$ and $n=0.95$ (with $25\%$ tensor contribution
to the CMB fluctuations on large scales).
The simulation box is $L=100h^{-1}$Mpc with $512^3$ fluid elements and
$256^3$ dark matter particles.
Three components are followed separately and simultaneously:
dark matter, gas and ``galaxies".
The last component is created continuously (like real galaxies)
during the simulation in places where local physical conditions
permit rapid cooling and collapse, the dynamics
of the aftermath of which cannot be followed in the present
simulation.
Instead, we allow star formation to occur in these regions
where gravitational collapse cannot be reversed
until stellar systems are formed, under plausible assumptions.
In addition to standard physics in cosmological
gasdynamic simulations,
feedback into the intergalactic medium (IGM)
from star formation is allowed in three related forms:
UV radiation, supernova energy and mass ejection.
Cooling due to metals is also included.
The model reproduces
the observed evolution of the luminosity density of the
universe at various energy bands (Nagamine \etal 1999),
the evolution of galaxy clustering (Cen \& Ostriker 1998)
and metallicity distributions (Cen \& Ostriker 1999b)
among others.

The results from this new simulation focusing
on the evolution of the cosmic gas have been presented 
in Cen \& Ostriker (1999a) and are summarized here.
We divide the baryonic gas into three temperature 
ranges (1) $T>10^7~$K (the X-ray emitting gas
in collapsed and virialized clusters of galaxies);
(2) $10^7~$K$>T>10^5~$K gas, 
which we will call the warm/hot gas and is located
outside of clusters of galaxies;
(3) $T<10^5~$K warm gas, 
which is seen in optical studies as $\lya$ clouds or Gunn-Peterson effect
and is primarily in voids at $z=0$.
A last component (4) is the cold gas that has been condensed
into stellar objects, which we designate ``galaxies".
Figure 1 shows the evolution of these four components,
and the results are consistent with our other knowledge.
Most of the baryonic mass is in warm gas (Lyman alpha forest)
at $z=3$ making up $94\%$,
which declines with increasing time to $26\%$ at $z=0$,
consistent with the HST observed clearing of the forest
and of low-z redshift $\lya$ cloud gas (e.g., Shull 1996, 1997).
The hot component increases in mass fraction with increasing time,
reaching $12\%$ at $z=0$,
and is consistent with observations of the local
properties of the X-ray emitting great clusters 
(e.g., White \etal 1993;
Cen \& Ostriker 1994; Lubin \etal 1996; Bryan \& Norman 1998).
The condensed component remains small,
consistent with the known mass density in galaxies (e.g., Fukugita \etal 1997).

\begin{figure}[t!] 
\centerline{\epsfig{file=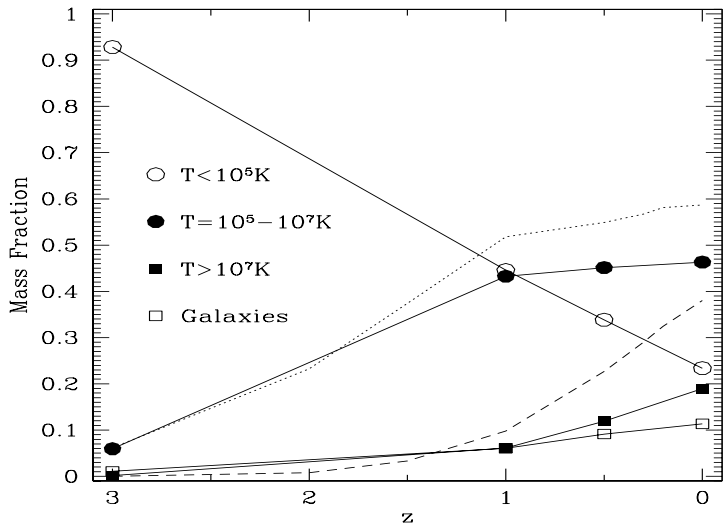,height=3.0in,width=4.0in}}
\vspace{0pt}
\caption{Mass Fractions of the four components (see text) as a function of redshift.}
\label{fig1}
\end{figure}

\begin{figure}[p] 
\centerline{\epsfig{file=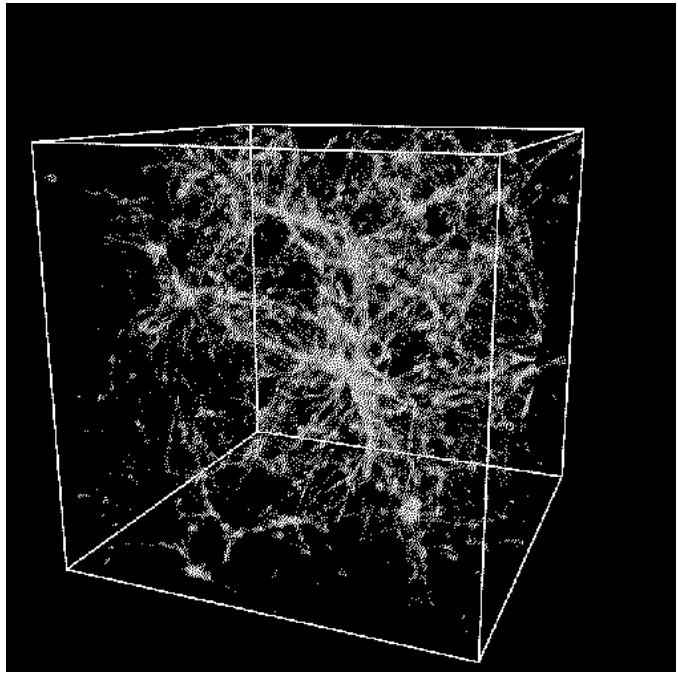,height=3.5in,width=3.5in}}
\vspace{10pt}
\caption{Spatial distribution of the ``warm/hot gas" in a box of size
$100h^{-1}$Mpc.}
\label{fig2}
\end{figure}

\begin{figure}[p] 
\centerline{\epsfig{file=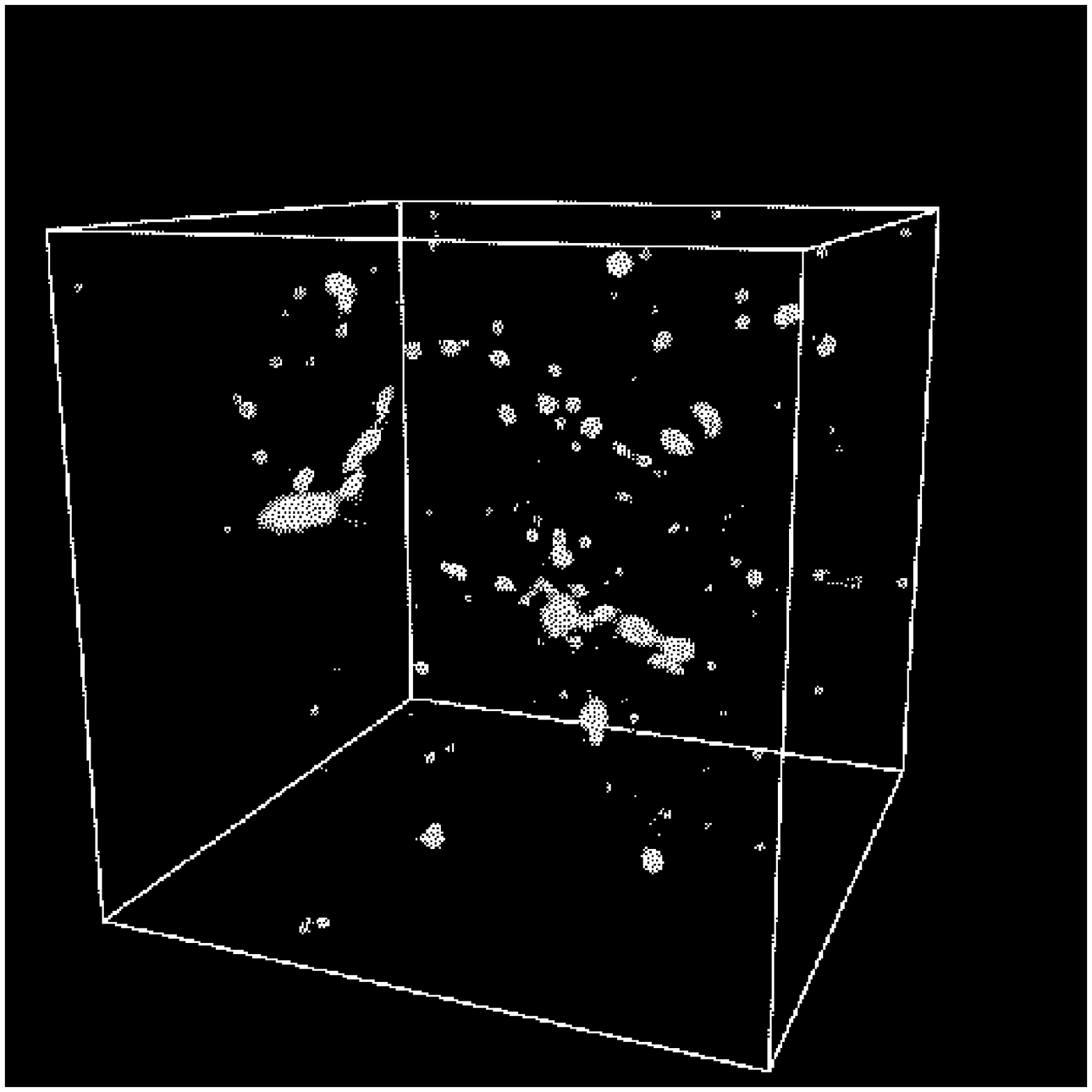, height=3.5in,width=3.5in}}
\vspace{10pt}
\caption{Spatial distribution of the hot intracluster gas ($T>10^7$Kelvin) 
in a box of size $100h^{-1}$Mpc.}
\label{fig3}
\end{figure}

Our attention will be focused on the solid circles in Figure 1:
the warm/hot gas rises rapidly
with increasing time and dominates
the baryonic mass budget by $z=0$,
reaching $52\%$ of the noncondensed mass fraction
or 47\% of the total baryons.
Also shown in Figure 1 is the warm/hot component for
two other models, an open CDM model
with $\Omega_0=0.40$ and $\sigma_8=0.75$ (dotted curves), and
a mixed hot and cold dark matter model with 
$\Omega_{hot}=0.30$ and $\sigma_8=0.67$ (dashed curves)
computed completely independently by Bryan \& Norman (1998).
Quite reassuringly, their results are in excellent agreement with ours.
The density fluctuation amplitude normalization of 
their mixed dark matter model 
is somewhat below that required to produce the abundance
of local galaxy clusters.
Therefore, an appropriately normalized mixed dark matter model
would yield a larger warm/hot gas fraction 
thus is in still better agreement with the other two models,
re-enforcing the conclusion that 
the warm/hot gas makes up most of the 
baryonic matter today, independent of models as long as
each model is normalized to match the local large-scale structure.
Physically, this may be understood as follows.
The temperature of the bulk of the gas today should
be determined, to the zeroth order,
by the velocity of converging waves that are collapsing today.
The length of the waves that have become nonlinear today
is about $8h^{-1}$Mpc, which is almost exactly the scale
that is used to normalize each model to match the local abundance
of clusters of galaxies.
It is worth stressing that the reason for most of the gas phase
being in the warm/hot gas is primarily gravitational, as implied above.
In other words, the gas is primarily shock heated 
during the gravitational collapse and formation of the 
present large-scale structure.
Other potentially relevant physical processes, such as
the meta-galactic radiation field,
metal cooling
and energy deposition into IGM from young galaxies,
which were included in the simulation examined here,
are shown to be not of primary importance (Cen \& Ostriker 1999a),
with increasing, secondary importance in that order.

Figure 2 shows the spatial 
distribution of this warm/hot gas (the box size is $100h^{-1}~$Mpc) at $z=0$.
Figure 3 shows the spatial 
distribution of hot cluster gas with $T>10^7~$Kelvin at $z=0$.
We see in Figure 2 a filamentary network 
of the warm/hot gas with
hot cluster gas (see Figure 3) residing at the intersections of 
filaments.
These regions of 
warm/hot gas - groups of galaxies, filaments and sheets -
typically have such a low surface brightness 
that current instrumentation does not detect
them as distinct ``sources".
The typical size (width of the filaments)
is about one to several megaparsecs, corresponding to 
an angle of order a degree or so, if placed
at a distance of a few hundred megaparsecs.
The lengths of these filaments are tens of megaparsecs.
The typical density of the filaments is about
$10-100$ times the mean density of the universe.

\section*{Contribution To The Soft X-ray Background From The Warm/Hot Gas}

It was shown 
in a simulation of a similar model but with somewhat lower 
resolution (Cen \etal 1995)
that this warm/hot gas 
makes a nontrivial contribution
to the soft X-ray background at $<1.0~$keV.
In the present higher resolution simulation
we confirm this conclusion.
After scaling $\Omega_{b}$ down from 
$0.049$ to $0.037$ using latest observations (Burles \& Tytler 1998)
and noting the $J\propto \Omega_{b}^2$ scaling relation,
we find that $25\%$ of the {\it total}
extragalactic soft X-ray background 
at $0.7~$keV comes from
this diffuse warm/hot gas: $J_{WH}=7~$keV/sec/cm$^2$/keV/sr.
Most of the emission at $<1.0$keV from the warm/hot gas
is due to various blends of emission lines.
Figure 4 shows 
X-ray emissivity of the background gas (i.e., warm/hot gas;
dashed cruve) 
and of all the bright clusters with $L_{bol}>10^{43}$erg/s (solid curve)
at redshift $z=0$.
The primary spectral signature of the background 
in the region 0.5-1.0keV
is the  ``iron bump" (a mixture of iron and oxygen lines, primarily).
Major lines in this range include
a blend of OVII lines at 0.561-0.574keV (characteristic of $10^6$~K gas),
a blend of lines from NeIX at 0.904-0.922keV (at a similar temperature),
OVIII line at 0.654keV (peaked around $10^{6.4}~$Kelvin; 
the ratio of the 0.57keV to the 0.65keV features can be used
as a temperature indicator)
and a pair of iron XVII lines at 0.726keV and 0.739keV 
(which have peak strength at approximately $10^{6.5}~$K).

\begin{figure}[t!] 
\centerline{\epsfig{file=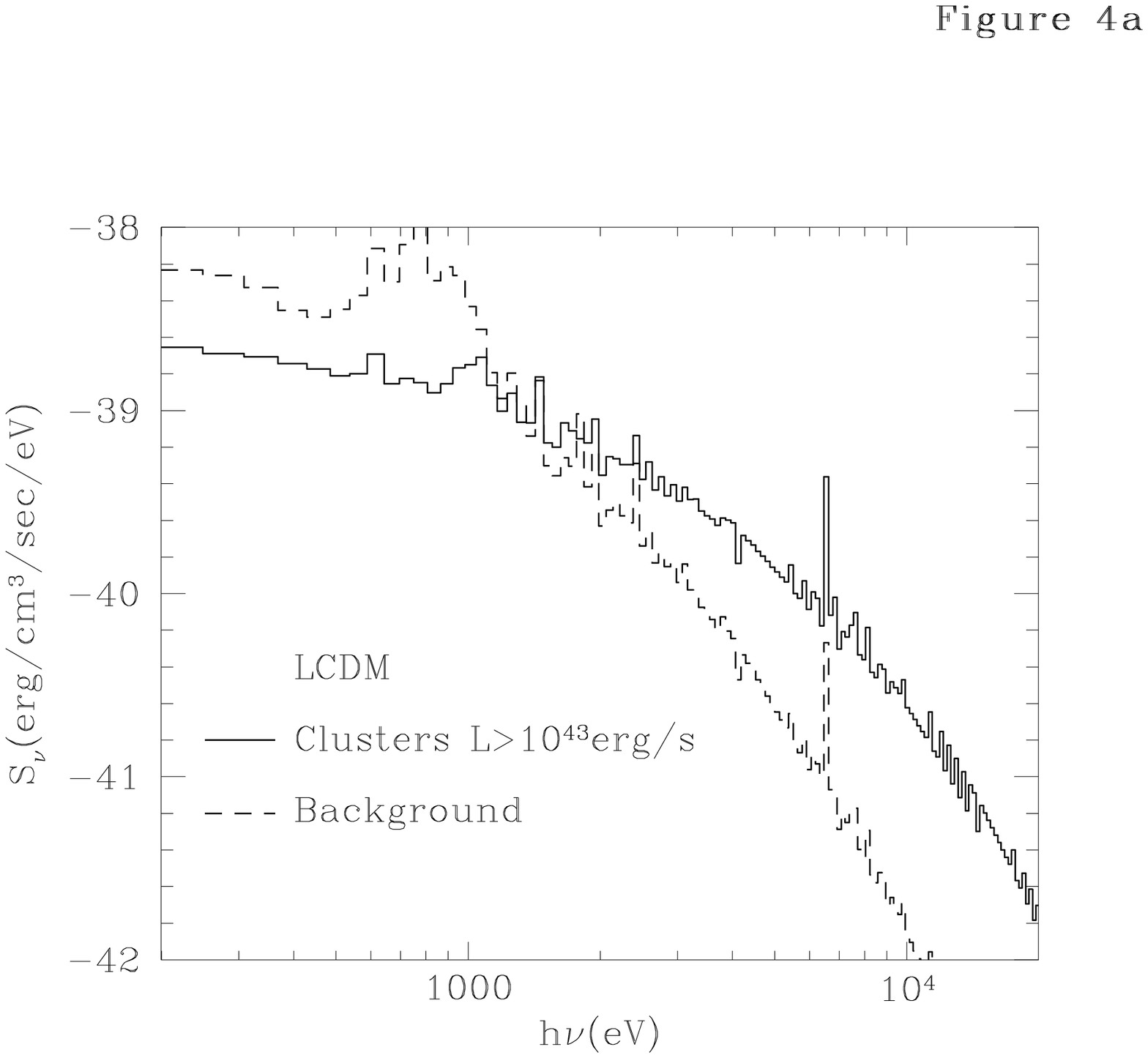,height=3.0in,width=5.0in, angle=270.0}}
\vspace{0pt}
\caption{Emission from warm/hot gas and hot intracluster cluster gas as
a function of frequency.}
\label{fig4}
\end{figure}

In the soft X-ray the Milky Way
is a strong source of emission at an effective temperature of $\sim 10^6~$K.
To some extent the soft X-ray 
background that we are proposing is the same as Galactic emission
in that the sum over all galaxies like our own
does make a nontrivial
contribution to the soft X-ray background,
since some of the warm/hot gas resides in the halos of galaxies.
If the average galaxy emits $2\times 10^{39}$erg/s in
the 0.5-1.5keV band (\cf Fabbiano 1989),
then the sum of all such will
produce an emissivity of $10^{-39}$erg/cm$^3$/sec/eV
not far from the levels shown in Figure 4.
But most of the emission shown by the dashed line in Figure 4
is due to hot gas in bigger systems than our own galactic disk or halo.
Thus, the temperature (as weighted by $\rho^2$) is in the range $10^6-10^7~$K,
whereas, for most of the galactic coronal gas,
the typical temperature
(as weighted by $\rho^2$)
is in the range $10^5-10^6~$K.
Thus, the line ratios indicative of the 
OVI/OVII and OVII/OVIII ratios will help in distinguishing
between the two components.
The spectral features which were noted above should help provide
clues to the origin of the background radiation in this range
as the galactic component is probably
too cool to produce
the iron blend which should be prominent in the
background component described here.
The steepening of the spectrum below 1keV
seen recently by ASCA (\cf Gendreau \etal 1995)
as well as the OVII lines (also noted by ASCA)
are also tentative but direct observational evidence
that the background gas we are discussing has already been detected.
Recent ROSAT observations of the soft X-ray background (Wang \& McCray 1993)
seem to hint the existence of this warm/hot gas.

Shadowing by nearby neutral hydrogen-rich
galaxies (McCammon \& Sanders 1990;
Wang \& Ye 1996; Barber, Roberts, \& Warwick 1996)
quite convincingly shows that 
a significant fraction of at least the component in the range 0.5-1.0keV
is truly extragalactic (\cf Burrow \& Kraft 1993;
Gendreau \etal 1995).
Wang \& Ye (1996) estimate that
$\ge 4~$keV/sec/cm$^2$/keV/sr 
out of the total soft X-ray background
at $0.7~$keV is truly diffuse in nature.
It thus deos not seem difficult for the contribution from the warm/hot gas
to account for the residual, diffuse soft X-ray background.

It is found from simulations that one half 
of the soft X-ray background due to the warm/hot gas
is emitted by structures at redshift $z<0.65$ and
three quarters from $z<1.0$. So one may be
be able to identify the optical features associated 
with the emitting gas.

\section*{Ways To Detect The Warm/Hot Gas}

\begin{figure}[t!] 
\centerline{\epsfig{file=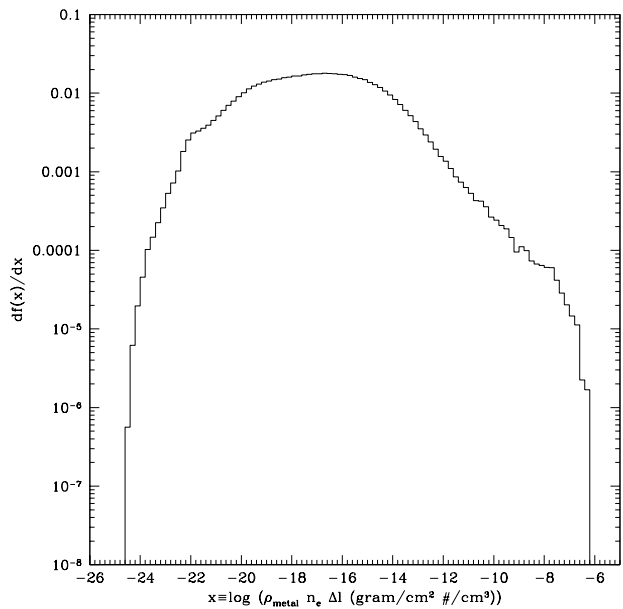, height=3.5in,width=3.5in}}
\vspace{10pt}
\caption{
shows the histogram of $x$ (defined in the abscissa),
which is proportional to the emission,
for gas in the temperature range $2\times 10^5-5\times 10^5~$K.
This temperature range is picked to best show
the abundance of O VI.
$\Delta l$ corresponding to $300$km/s Hubble velocity 
is used in the calculation. }
\label{fig5}
\end{figure}

The spectral features of this warm/hot gas
are in the EUV
and soft X-ray,  
which make it difficult to observe at low redshift
due to possible confusion with features from 
the interstellar medium in our own galaxy.
Proper identifications of
spectral lines will allow to unambiguously distinguish local galactic
features from extragalactic ones, for sufficiently
distant, {\it individual filamentary} structures (perhaps $z\ge 0.01$).

These individual structures
may be detected directly in several ways.
First, Hellsten \etal (1998), on the basis of similar  simulations,
have predicted the existence of an X-ray absorption forest due to
ionized oxygen (O VII 574~eV line)
in the warm/hot temperature range.
Perna \& Loeb (1998) have also made similar calculations based on
simplified models for the IGM.
Work underway with T. Tripp and E. Jenkins indicates
that UV absorption lines (OVI 1032A,1038A doublets)
due to gas in this temperature range
primarily in the distant outskirts of galaxies
may be detectable by current or planned instruments.
Second, strong soft X-ray emission lines from highly ionized species
(such as O VIII 653~eV line)
should also be observable (Jahoda \etal 1998, in preparation).
Figure 5 shows a histogram of emission 
focusing on the warm/hot gas in a narrow range of temperature relevant for OVI.
Future UV observations 
such as proposed PRISM, HIMS or HUBE SMEX missions
may be able to detect the gas at the high end of $x$.
Therefore, the next generation of UV to soft X-ray spectroscopic instruments
with sufficient spectral resolution and sensitivity
will provide direct ways to measure 
``{\it the X-ray forest}" or ``{\it the X-ray large-scale structure}",
depending on whether it is seen in absorption or emission.
Third, the warm/hot gas may show up as very broad, relatively
weak (mostly having $N_{HI}\le 10^{13}$cm$^{-2}$ with a 
small fraction at higher column densities),
low redshift $\lya$ clouds (Shull 1996,1997;
but observations more sensitive than current ones
are required). NGST should be able to identify
these local counterparts of the high redshift $\lya$ clouds.

If our prediction that most of the residual soft X-ray background
is produced by the warm/hot gas is correct,
then, since most of the soft X-ray background is produced
by low redshift structures,
this could be verified by associations of this soft X-ray
background radiation field
with relatively nearby large scale structure features,
as well as by the soft X-ray angular auto-correlation function.
A preliminary study (Cen \etal 1999, in preparation)
indicates that the auto-correlation function
of the soft X-ray background from warm/hot gas
is positive up to a few degrees in separation,
in good agreement with observations from
ROSAT (Soltan \etal 1996).
Cross-correlating soft X-ray background (Refregier, Helfand, \& McMahon 1997)
or Sunyaev-Zel'dovich effect (Refregier, Spergel, \& Herbig 1998) with
galaxies should provide additional important information.
The next generation of X-ray instruments (AXAF and ABRIXAS),
in conjunction with large-scale redshit surveys (e.g., Sloan Digital
Sky Survey) should provide some potentially powerful tests of 
the existence of this warm/hot gas.

\section*{Conclusions}

According to the current theory for the growth of cosmic structure,
it is inevitable that most of the cosmic gas is shock
heated during the course of gravitational
collapse and formation of the present-day large-scale structure
and ends up in a warm/hot gas with a
temperature $10^5-10^7~$Kelvin in the present-day universe.
This gas accounts for the so called missing
baryons in our local universe, which constitutes most of
the baryons today.
This gas with density of $10-100$ times the mean density of the universe
resides in a filamentary network with each individual
filament having a length of tens of megaparsecs and a width
of order one megaparsec,
easily distinguishable from the much
hotter gas in the centers of great clusters of galaxies,
which is located in the intersections of the filaments
and has a much higher temperature.
The emission (primarily line emission)
from this warm/hot gas dominates
that from intracluster gas in the soft X-rays and 
may account for most of the 
residual soft X-ray background at $<1$keV.
This network of filamentary warm/hot gas
is probably too faint to be detected as individual sources
by available instruments.
However, the next generation of instruments in EUV and soft X-ray
should be able to detect this warm/hot gas,
in a number of possible ways, suggested in the previous section.
If this warm/hot gas is detected, a consistent
picture of gravitational growth of structure
will be affirmed once again;
otherwise, the current theory of structure formation 
within the gravitational instability paradigm
and/or of standard
light element nucleosynthesis may require a re-examination.

\vskip 1.cm
The work presented here should be mostly
credited to my collaborators, Drs. Jeremiah P. Ostriker,
Hyesung Kang and Dongus Ryu.
I thank Dr. Greg Bryan for allowing to use the results
from his simulations before publication.
I thank Dr. Steven Brumby for his warm hospitality and
for organizing an entertaining and stimulating conference.
The work is supported in part by grants AST9318185 and ASC9740300.


\end{document}